\documentclass[letterpaper]{article} %
\usepackage{aaai23}  %
\usepackage{times}  %
\usepackage{helvet}  %
\usepackage{courier}  %
\usepackage[hyphens]{url}  %
\usepackage{graphicx} %
\usepackage{natbib}  %
\usepackage{caption} %
\usepackage{newfloat}
\usepackage{listings}
\DeclareCaptionStyle{ruled}{labelfont=normalfont,labelsep=colon,strut=off} %
\usepackage{booktabs, longtable, adjustbox, rotating}
\usepackage{tikz, bm, amsmath}	
\usepackage{subcaption}
\usepackage{amssymb}
\usepackage{multirow, tabularx}

\DeclareMathOperator{\E}{\mathbb{E}}

\title{Effects of Algorithmic Trend Promotion:\\ Evidence from Coordinated Campaigns in Twitter's Trending Topics}

\author{
Joseph Schlessinger,\textsuperscript{\rm 1}
Kiran Garimella,\textsuperscript{\rm 2}\footnote{Corresponding author}
Maurice Jakesch,\textsuperscript{\rm 3}
Dean Eckles\textsuperscript{\rm 4}
\\
}
\affiliations{
    \textsuperscript{\rm 1}U.S. Army\\
    \textsuperscript{\rm 2}Rutgers University\\
    \textsuperscript{\rm 3}Cornell University\\
    \textsuperscript{\rm 4}MIT\\
     joe.c.schlessinger@gmail.com, kiran.garimella@rutgers.edu, mpj32@cornell.edu, 
     eckles@mit.edu
}

\begin{document}
\maketitle

\begin{abstract}

In addition to more personalized content feeds, some leading social media platforms give a prominent role to content that is more widely popular. On Twitter, ``trending topics'' identify popular topics of conversation on the platform, thereby promoting popular content which users might not have otherwise seen through their network.
Hence, ``trending topics'' potentially play important roles in influencing the topics users engage with on a particular day.
Using two carefully constructed data sets from India and Turkey, we study the effects of a hashtag appearing on the trending topics page on the number of tweets produced with that hashtag.
We specifically aim to answer the question: How many new tweeting using that hashtag appear \textit{because} a hashtag is labeled as trending?
We distinguish the effects of the trending topics page from network exposure and find there is a statistically significant, but modest, return to a hashtag being featured on trending topics.
Analysis of the types of users impacted by trending topics shows that the feature helps less popular and new users to discover and spread content outside their network, which they otherwise might not have been able to do.
\end{abstract}

\section{Introduction}

Leading social media platforms feature personalized feeds of content, often largely based on what other accounts a user is connected to --- whether through directed or undirected ties or through group co-membership. However, some of these platforms also give a prominent role to content that is more widely popular.
``Trending topics'' is a  feature on Twitter that directs users to popular content on the platform within countries or regions.
The algorithms behind the feature identify viral content on the platform and provides users a way to access the content even if they are not following any user who tweeted about it.
Ever since its launch over a decade ago, according to Twitter, the feature has helped Twitter users ``understand what is happening in the world and what people's opinions are about it''~\cite{twitterTwitterTrendsFAQs}.
The feature may also allow Twitter to increase user engagement on the platform, by showing content to users that they would not have otherwise seen, and boost revenue through sponsored trends~\cite{trendingSponsored}.

The trending topics page can also be thought of as an agenda setting tool. It is a way to influence what people talk about:
Twitter implicitly endorses trending content as important and topical~\cite{proferes2019algorithms}.
The feature has been used in a wide range of high profile events, spreading messages and spurring conversations during crises events like earthquakes~\cite{sakaki2010earthquake}, protests~\cite{tufekci2017twitter,bruns2013arab}, or popular sporting events~\cite{knapp2013super}.
Trending topics have also been used in harassment campaigns~\cite{flores2017tweeting,chatzakou2017hate}, to spread conspiracies~\cite{dickson2019jeffrey} and disinformation~\cite{hindman2018disinformation}, and to amplify extremism~\cite{brooking2016atlantic}.
The trending topics page is a coveted spot for those seeking to promote a particular hashtag, as it provides algorithmic amplification to the content.
There are many examples of people manipulating the trending topics page to gain free advertising for their cause~\cite{jakeschTrendAlertHow,elmasLateralAstroturfingAttacks2019}.
All users identified as being in a particular geographical region see the same trending topics.
Once a topic is featured on the page, it is pushed by Twitter to millions of Twitter users in the region, potentially garnering more popularity.

Despite the important role the feature plays in popular discourse on social media, there has not been any study to date investigating the impact of a hashtag appearing on the trending topics page (henceforth just referred to as trending).
In this paper, we try to understand the causal impact of trending on tweet volume for a given hashtag. We answer the question: How many new tweets are \textit{caused} by a hashtag appearing on the trending topics page?
The analysis helps us understand the power of the trending topics page as an agenda setting tool; specifically, it explores the extent to which Twitter contributes to a hashtags' popularity by labeling it as trending.
Understanding this impact also offers insights into Twitter's control over its platform; can it simply pull the trending lever and make users talk about certain topics?

While the impact of trending is important to understand, its measurement is elusive for a few reasons.
First, there are multiple ways a user can be exposed to content on Twitter and it is difficult to know how a user has been exposed with data publicly available.
For example, users may engage with content because they saw it on the trending topics page, saw a friend tweet about it, or found it through search.
Second, events on Twitter can interact with the real-world or other social media sites, which further confounds efforts to isolate the effect of trending.
If a topic trends because of a related real-world event, there is no way to know whether engagement is caused by the actual event or its trending status.
Moreover, the exact algorithm behind trending topics is opaque and under constant revision~\cite{twitter2020trendingchanges}, making it difficult to understand why a specific hashtag is trending.
In order to characterize the specific impact of the trending topics page, one must account for all factors that can explain user engagement with a hashtag.
Under most circumstances, it is not possible to collect the necessary data to perform this analysis without special access from Twitter.

Our analysis relies on two natural experiments to quantify the return to trending.
We use datasets from two different contexts (Turkey and India) where astroturfed campaigns artificially promoted trends.
The inauthentic nature of these trends allows us to control for external events and thereby isolate the causal effect of trending.

We find that a hashtag trending causes a 60 to 130\% increase in new tweets within 5 minutes of being trending.
Even though this might appear high, the effect is modest in absolute terms, amounting to approximately one new tweet per minute.
We find that the effect of a hashtag trending varies widely across hashtags, which suggests content has some endogenous virality: certain trends were more engaging than others.
By measuring who gets exposed to the hashtag via the trending topics page, we demonstrate that the trending topics page allows a campaign to expose new parts of the network to the hashtag.
Thus, even though the trending topics page might have a modest absolute impact in terms of new tweets, it helps promote content to parts of the network that would not have been exposed otherwise.
These results have implications on how trending topics should be understood and the impacts of manipulating trending topics.

\section{Related Work}

Trending topics were introduced in 2010 as a tool to organize user-generated hashtags and help answer the ``What's happening?" question~\cite{burgess2020twitter}.
Twitter trends have been a hallmark feature ever since and have been constantly evolving. For instance, in September 2020, Twitter added more context to trends along with displaying representative tweets for a trend.
Trending topics serve an important purpose: exposing users to popular, viral and serendipitous content, which they might not otherwise not encounter through their network.
The exact algorithm behind identifying trending topics is not known; however, Twitter reveals that trending topics are created based  on ``both volume of a term but also the diversity of people and tweets about a term and looking for organic volume increases above the norm''~\cite{koumchatzky2017using}.

The trending topics feature is not unique to Twitter.
Facebook had a similar feature showing popular content on the platform in 2014 but discontinued it in 2018.\footnote{\url{https://about.fb.com/news/2018/06/removing-trending}}
Instagram and TikTok surface new, popular content, but recommendations are linked to users’ previous activity.
This is a relevant difference from Twitter's trends, which are mostly geographical; thus, everyone within the same region might see the same trends.
Trending topics are also a source of revenue for Twitter. Twitter offers ``promoted trends,'' where companies can pay to advertise a specific hashtag at the top of the trending topics page.\footnote{\url{https://business.twitter.com/en/advertising/takeover/promoted-trend.html}}

Twitter trending topics are regularly mentioned in mainstream news sources.\footnote{For instance, a search on Google news for the term `trending on twitter' yields hundreds of results from mainstream news sources in just the past week. \url{https://news.google.com/search?q=trending on twitter}.}
Other work has posited that Twitter trending topics have had an impact in encouraging democratic protests~\cite{tufekci2017twitter}, helping disaster response~\cite{ashktorab2014tweedr}, and enabling voices for the marginalized~\cite{milanMobilizingTimesSocial2015,jackson2016re}.
However, Twitter trends have also been manipulated and misused for nefarious purposes, such as the spread of conspiracies~\cite{dickson2019jeffrey}, misinformation~\cite{hindman2018disinformation}, hate speech~\cite{flores2017tweeting,chatzakou2017hate} and election manipulation~\cite{bail2020assessing}.
Elmas et al.~\cite{elmasLateralAstroturfingAttacks2019} estimate that these ``ephemeral astroturfing attacks'' are responsible for at least 20\% of top 10 global trends and up to 50\% of trends in Turkey. 
Manipulation of Twitter trends has been commonplace and well documented in research~\cite{zhangTwitterTrendsManipulation2017}, and journalistic pieces.\footnote{\url{https://qz.com/africa/2086139/twitter-has-suspended-its-trends-section-in-ethiopia/}}

Even though most of these cases (both positive and negative) are well studied, it is extremely hard to study the impact that Twitter trends had on real world outcomes.
Regarding elections, Twitter often echoes other media forms, rather than influence them. 
Candidates may use Twitter to generate publicity, but this ``buzz'' may be irrelevant to election outcomes \cite{gayo2012no,murthyTwitterElectionsAre2015}.
Due to lack of access to exposure data, which only the social media platforms have~\cite{lazer2020studying}, unless in special cases, it is not possible to make such inferences at scale.
\citet{asurTrendsSocialMedia2011} look at the emergence of trending topics of Twitter and the factors that influence them. They find that trends appearing on the trending topics page demonstrate linearly increasing engagement, but it is unclear whether this effect is \textit{caused} by the trending topics page.
At least some of the engagement driven by the trending topics is ``hashjacking'' where users capitalize on the trending status of a hashtag to advertise something unrelated \cite{dfrlabSouthAfricanTwitter2020}. 
While trending hashtags are well-researched, there is little work on the actual impact of the trending topics page.

There has been significant research on trying to identify and quantify influence of content posted on Twitter~\cite{riquelmeMeasuringUserInfluence2016}.
The most basic analysis looks at actions naively attributed to tweets, such as retweets, favorites, and replies, or look at the influence of users, based on their followers~\cite{chaMeasuringUserInfluence2010,bakshyEveryoneInfluencerQuantifying2011}.
\citet{romeroDifferencesMechanicsInformation2011} examine the relationship between influence and the number of exposures to a certain hashtag and show that content differs in the probability of adoption based on non-zero exposure, but it also differs in the extent to which additional exposures increase probability of adoption.
Despite much effort, influence may be impossible to quantify in a meaningful way, especially with an observational study. 
Homophily complicates attempts to quantify peer effects with observational data \cite{bakshyRoleSocialNetworks2012}. 
Moreover, unobservable factors outside of the network may be responsible for user behavior. In one estimate, 29\% of information comes from factors beyond Twitter \cite{myersInformationDiffusionExternal2012}.

Our work falls into the category of studies performing algorithmic audits on socio-technical systems.
The two typical approaches used in literature for audits are to make use of bot accounts or to work in collaboration with the platforms.
Many socio-technical platforms have been audited in the past couple of years, including Google~\cite{robertson2018auditing}, Uber~\cite{chen2015peeking}, Amazon~\cite{juneja2021auditing}, and YouTube~\cite{ribeiro2020auditing}.
Specifically on Twitter, recent studies \cite{bandy2021more,bartley2021auditing} perform an audit on the algorithmic ranking feature introduced by Twitter in 2017.
Using bot accounts, the authors study the properties of the algorithmic timeline in terms of the types of content it surfaces, when compared to the chronological timeline.
Another recent study~\cite{huszar2022algorithmic} performs a large scale audit of algorithmic amplification of politics on Twitter and finds that in six out of seven countries studied, the mainstream political right enjoys higher algorithmic amplification than the mainstream political left. They also find that right leaning media sources are also favored by Twitter's algorithms.
Unlike previous approaches in this realm, our approach makes use of a natural experiment to study the impacts of trending topics on Twitter.

\section{Data}

One of the contributions of our paper is the curation of datasets which allows us to study the causal impact of trending topics. We make use of two large-scale datasets from Turkey and India containing hashtags which were artificially trended using different astroturfing techniques. A quantitative summary of the datasets is shown in Table~\ref{tab:datasets}.

\subsection{Case Study 1: Turkey}

The first dataset comes from \citet{elmasLateralAstroturfingAttacks2019}.
The dataset consists of hundreds of hashtags from Turkey which were manipulated using a technique termed ``ephemeral astroturfing''.
The attack takes place as follows: attackers control tens of thousands of compromised accounts of real individuals, which are used to tweet a specific hashtag in unison.
These tweets only stay active for a few seconds before they are deleted. 
However, due to a sudden surge in the tweet volume, Twitter labels the hashtag as trending. 
As a result, these hashtags appear on the trending topics page largely without associated tweets, leading other users to add their own content using the hashtags.
The tweets for this dataset were collected in two stages. 
First, Elmas et al. collected 1\% of all real-time tweets using the Twitter streaming API and discovered astroturfed trends by detecting the unique properties of the campaigns. 
They also collected complete global and Turkish trending information at a 5 minute interval. 
Second, using the Twitter search API, we augmented this data by collecting all existing tweets containing the astroturfed hashtags, along with the full follower networks of participating users. 
In this paper, we made use of the hashtags which reached the top 50 in Turkish trends at some point in time in July 2019, which amounts to 418 hashtags with a total of over 3 million tweets made by over 790,000 users.

Most of the manipulated hashtags were illicit advertisements (e.g., gambling sites), politics, policy appeals, and boycotts.
These are unique because the trends are promoted with tweets that use nonsensical assortments of Turkish words (e.g. ``to organize milk frost deposit panel'').
They are deleted soon after they are posted, so the trend appears on the Trending Topics page with few associated tweets. 

\subsection{Case Study 2: India}

The second dataset was collected during the 2019 general elections in India by \citet{jakeschTrendAlertHow}.
One of the leading political parties in the election (the BJP) created an infrastructure of social media accounts for online campaigning and mobilization of voters. 
The infrastructure consisted of a network of volunteers incubated through public WhatsApp groups. 
Through these volunteer networks, organizers coordinated ``trend attacks'' on Twitter.
First, through the WhatsApp groups, leaders disseminated links to Google Docs that contained a bank of a few hundred ``template'' tweets (example tweets) for a hashtag, that individuals could choose from.
At a predetermined time, the volunteers copy pasted these template tweets from their personal Twitter accounts.
The coordinated action would send a particular hashtag to the trending topics page by tweeting in unison, thus fooling the Twitter trending algorithm into thinking the hashtags were popular.
For a more detailed explanation of the dataset and the manipulation, see \citet{jakeschTrendAlertHow}.

The dataset contains 75 such campaigns, each with a corresponding hashtag, making up over 2.3 million tweets from 244,000 users. 
The campaigns were seemingly quite successful at their stated goal of making hashtags trend; 62 of the 75 campaigns were trending across India on the day the campaign was organized. 
For this dataset, we only have half-hourly trending data indicating whether a hashtag was trending nationally every 30 minutes.
We also obtained the follower and friend network of all the users in our dataset.

The manipulated hashtags are political messages, but not necessarily distinct from ``natural'' political trends. They mostly support events or initiatives of the BJP. 
They differ from regular trends in that they are due organizers coordinating a volunteer network in the background to post pre-written messages at an agreed time.

\begin{table}[]
\centering
\caption{Summary of the two datasets.}%
\label{tab:datasets}
\begin{tabular}{ccccc}
\hline
Dataset & Hashtags & Tweets & Users & \begin{tabular}[c]{@{}c@{}} Median \\tweets/hashtag\end{tabular} \\ \hline
Turkey  &  418        & 3M       & 790k             & 693 \\ %
India   &  75         & 2.36M         &  244k               & 5268 \\ %
\hline
\end{tabular}
\end{table}

\subsection{Importance of dataset choice}
Both our datasets were carefully curated to fit the objective of our study. 
One of the main obstacles to disentangling the effects of the trending topics page is the presence of external events.
In both case studies, we know that the investigated trends have been artificially created. 
Moreover, we know which tweets were used to artificially promote the hashtag. 
If we account for these artificial tweets, we are left with tweets that are genuinely engaging with the hashtag \textit{and} are not responding to an event outside of Twitter.
We can then use information about the friend network (who the users follow) to separate the effect of the trending topics page from natural diffusion across the network.
Finally, the dataset is linked to trending information.
For all investigated trends, we know when the hashtags appeared on the trending page within some margin of error.
With the Turkish data, we know the trending time within 5 minutes and with the India data, we know the trending time within 30 minutes of uncertainty.
Data on when hashtags appear on the trending page are not available through Twitter's API.

\subsection{Terminology}

Our methodology assumes hashtag use can be explained by three proximate causes: (i) a users adopts the hashtag after someone in their friend network uses it, (ii) from the trending topics page, or, or (iii) because of some external event.\footnote{Admittedly, this is a simplification.
These factors need not be independent. 
For example, a user may be thinking about using a hashtag because someone they follow uses it. 
Then, they see the hashtag on the trending topics page and decide to tweet. 
On top of that, users can be exposed to tweets from users outside their friends network, such as through Twitter's algorithmically-curated timeline.
Please see Section~\ref{sec:discussion} for discussion on how these factors might impact our findings.}
Through out the rest of the paper, we refer to two sets of users/tweets based on how they got exposed to a hashtag prior to themselves posting the hashtag.
If a user tweets a hashtag after one of their friends tweet with the same hashtag, we call the user \emph{network exposed}, indicating that they were likely exposed through their friend network.\footnote{By extension we label the tweets by such users as network exposed tweets.}
On the other hand, if a user tweets a hashtag without any prior exposure through their friend network, we assume that they saw the hashtag through the trending topics page and label the user \emph{trending exposed}.

\begin{figure}
    \centering
    \includegraphics[width=.5\textwidth]{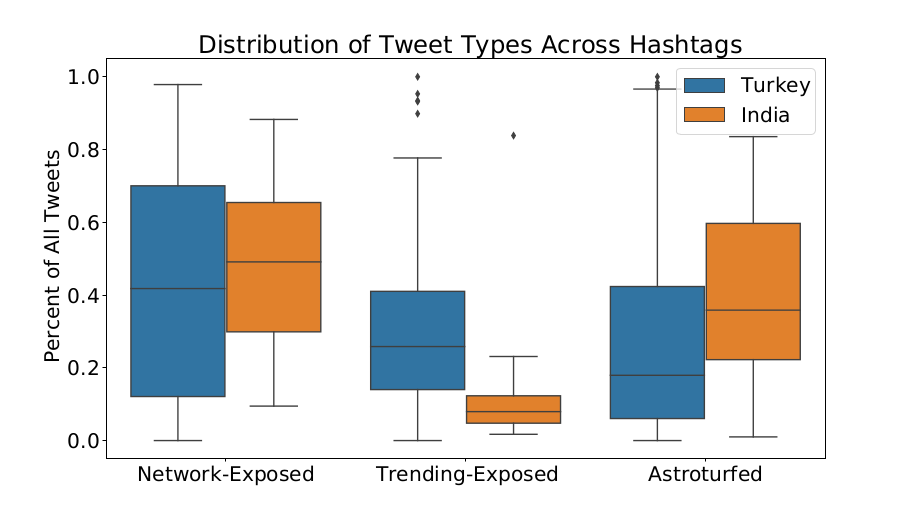}
    \caption{Distribution of network, trending exposed, and astroturfed tweets in our datasets. Each observation is a single hashtag.} %
    \label{fig:tweet_exposure_statistics}
\end{figure}

Figure~\ref{fig:tweet_exposure_statistics} shows the fraction of network, trending exposed, and astroturfed tweets in our dataset. We can see that the Turkish dataset has a much higher fraction of exposure through the trending topics page. This is expected given the way the manipulation works, where most original tweets that get the hashtag trending are deleted, thus leaving a significant fraction of the exposure through the trending topics page.

\section{Causal Impact of Twitter's Trending Topics} \label{returnstotrending}
In this section, we study the causal impact of a hashtag trending on hashtag tweet volume. 
This is typically a challenging query because, for any hashtag, we only observe the trending and network exposed factors, but the user may already be aware of the hashtag because they heard of it through an external source (e.g., it being a popular topic that day). 
Such external, often unobserved events, make it difficult to measure returns to trending.
Hashtags often have a real-world analogous event that drives hashtag engagement.
Using the Super Bowl as an example, many tweets will use \#SuperBowl, but it may be impossible to determine \textit{why} they are using it. 
It could be because of tweets from their Twitter friends, the trending topics page or because of the event itself.

In our carefully chosen datasets, we observe relevant external events. 
These hashtags were created and promoted by a group of actors for the sole purpose of trending and did not have any prominent external events associated with them prior. 
For instance, in the case of Indian dataset, we have the messages from the WhatsApp groups in which the participants plan the tweets and can identify template tweets (from lists of tweets in documents shared in the WhatsApp groups).
In the case of the Turkish data, the tweets used to create the trend are deleted almost instantly \textit{and} they are semantically distinct from normal tweets. 
Thus, we can more safely assume that there is no other contemporaneous external event that impacts the tweet volume. 
If we remove the tweets we know are part of the manipulation (i.e. the astroturfed tweets), we arrive at the following model to estimate the causal effect that the trending topics page has on new tweets produced:
\begin{equation}
	\log(\E(Y_{th})) = \alpha + \lambda D_{th} + \tau t + \gamma (D_{th} \times t) + \beta E_{th} + \xi_h,
	\label{eq:trending_causal_model}
\end{equation}
where $Y_{th}$ is the number of tweets with hashtag $h$ at time $t$, $D_{th}$ is a binary indicator for whether the hashtag $h$ is trending at time $t$, $E_{th}$ is the tweets that came after friends network exposure, and $\xi_h$ are hashtag-specific fixed effects. 
Here $\lambda$ is the causal effect of interest, which represents the volume of new tweets that happen specifically because of the hashtag appearing on the trending topics page.
Additionally, $\gamma$ is the interaction effect between the trending and time, and in practice represents the trending effect over time.
In effect, we are measuring the increase in tweets from users who had no exposure to the hashtag prior to posting (trending exposed tweets) after the hashtag appears on the trending topics page.
It is important to note that we are estimating the \textit{immediate} impact of a hashtag appearing on the trending topics page. 
Near the initial time of trending, there is very little exposure through the network.
However, as time goes on and more tweets occur, much of the network is exposed.

\begin{figure}[ht]
	\centering
	\includegraphics[width=.5\textwidth]{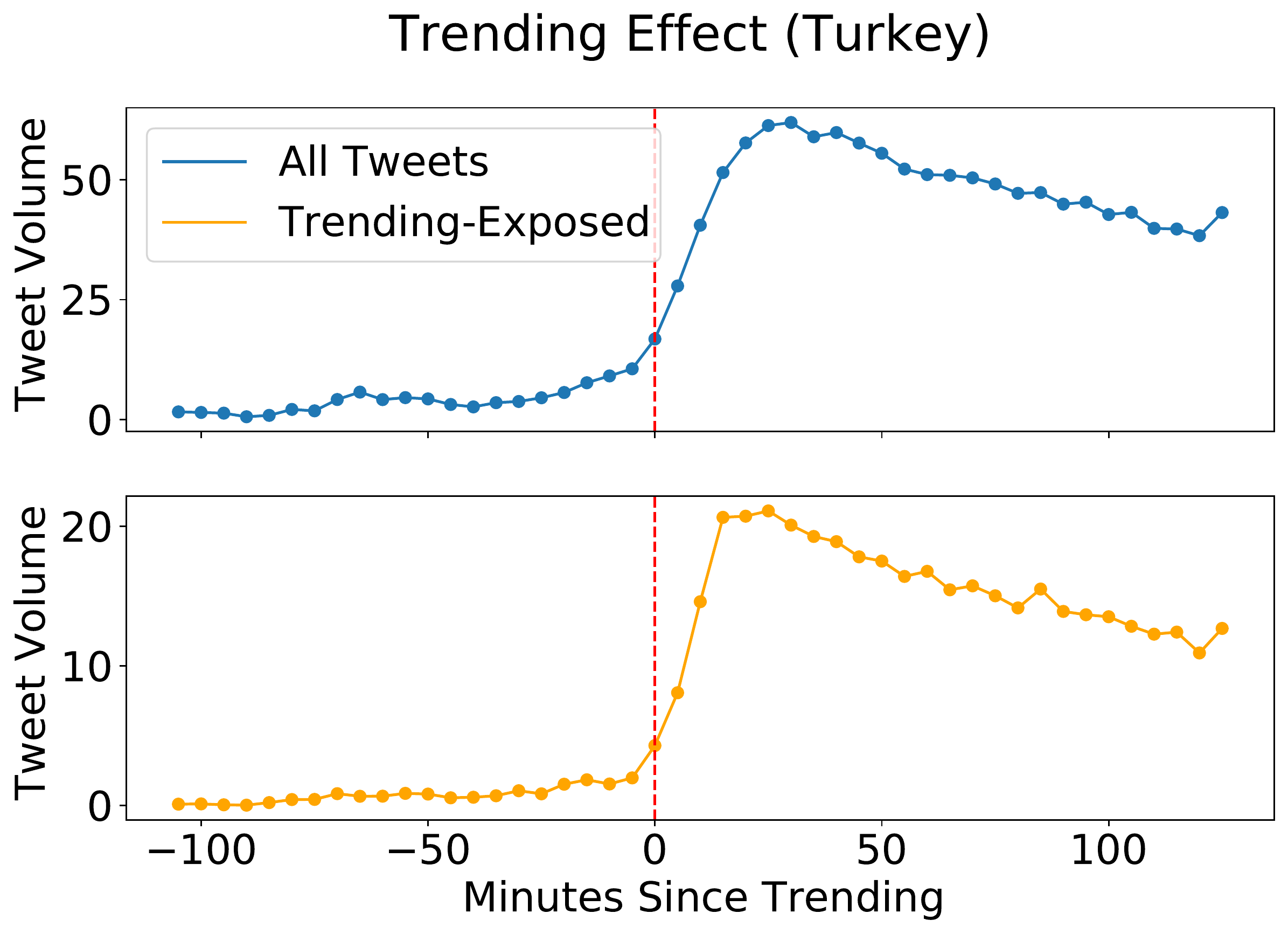}
	\caption[Turkey Event Study]{The average number of tweets (per five minutes) for all hashtags over time in the Turkish dataset. The top panel shows the average across all the tweets, while the bottom panel shows on only the tweets that we estimate to be potentially exposed through the trending topics page. Time is adjusted to be minutes from appearing on the trending topics page. Note the sudden jump in the tweet volume right at the time the hashtag was trending.}
	\label{fig:event_studies_turkey}
\end{figure}

Figure~\ref{fig:event_studies_turkey} shows a visual depiction of our setup in the Turkish dataset.  The figure depicts the average tweet volume over all hashtags in our Turkey dataset over time.
The top part of the Figure includes all tweets, including the exposures through the network and the trending topics page, while the bottom specifically focuses on the tweets we estimate to have come through trending exposure.
The x-axis shows the time in minutes from the hashtag appearing on the trending topics page.
We can see a clear increase in the mean tweet volume at time 0, which is when the hashtag starts trending.
We use the increase in trending exposed tweets, i.e., tweets that we assume come from exposure via the trending topics page, to estimate the causal effect.

\subsection{Modeling Considerations} 
We estimate Equation \ref{eq:trending_causal_model} using a quasi-Poisson generalized linear model, which uses a log-link but which does not impose the assumption of equal mean and variance of the Poisson distribution.
We use a Poisson model because our outcome of interest is a count of the number of tweets using a hashtag binned every five minutes.
Thus, the coefficients we report are a the logarithm of the multiplicative increase in new tweets after the hashtag started trending.
In all the results, we use clustered standard errors at the hashtag level~\cite{liang1986longitudinal}.
All the regression models reported included hashtag-specific fixed effects.

While we collected trending information for the hashtags in both our datasets, we did so at different temporal granularities.
For the Turkish dataset, we collected whether a hashtag was trending every 5 minutes (and thus have only a 5 minute error interval). On the other hand, for the Indian dataset, we only have hourly trending information, which means our trending times could be off by an hour in the worst case.
Thus, we have an hour long period of uncertainty about the treatment time or when the hashtag appeared on the trending topics page. 
To deal with this period of uncertainty in the Indian dataset, we run one set of models assuming the hashtag trended at the beginning of this period (\emph{Earliest} model) and one model where we remove all data from the period of uncertainty (\emph{Donut hole} model). The latter strategy is commonly used in the regression discontinuity literature and is referred to as the ``donut hole'' approach~\cite{cattaneo2021regression}.
For instance, if we record that a hashtag was trending during the period 10--11 am, our \emph{Earliest} strategy would assume that the hashtag started trending at 10 am, whereas the \emph{Donut hole} approach ignores all data during the hour and considers data starting at 11 am.

\begin{figure}[h]
	\centering
	\includegraphics[width=.5\textwidth]{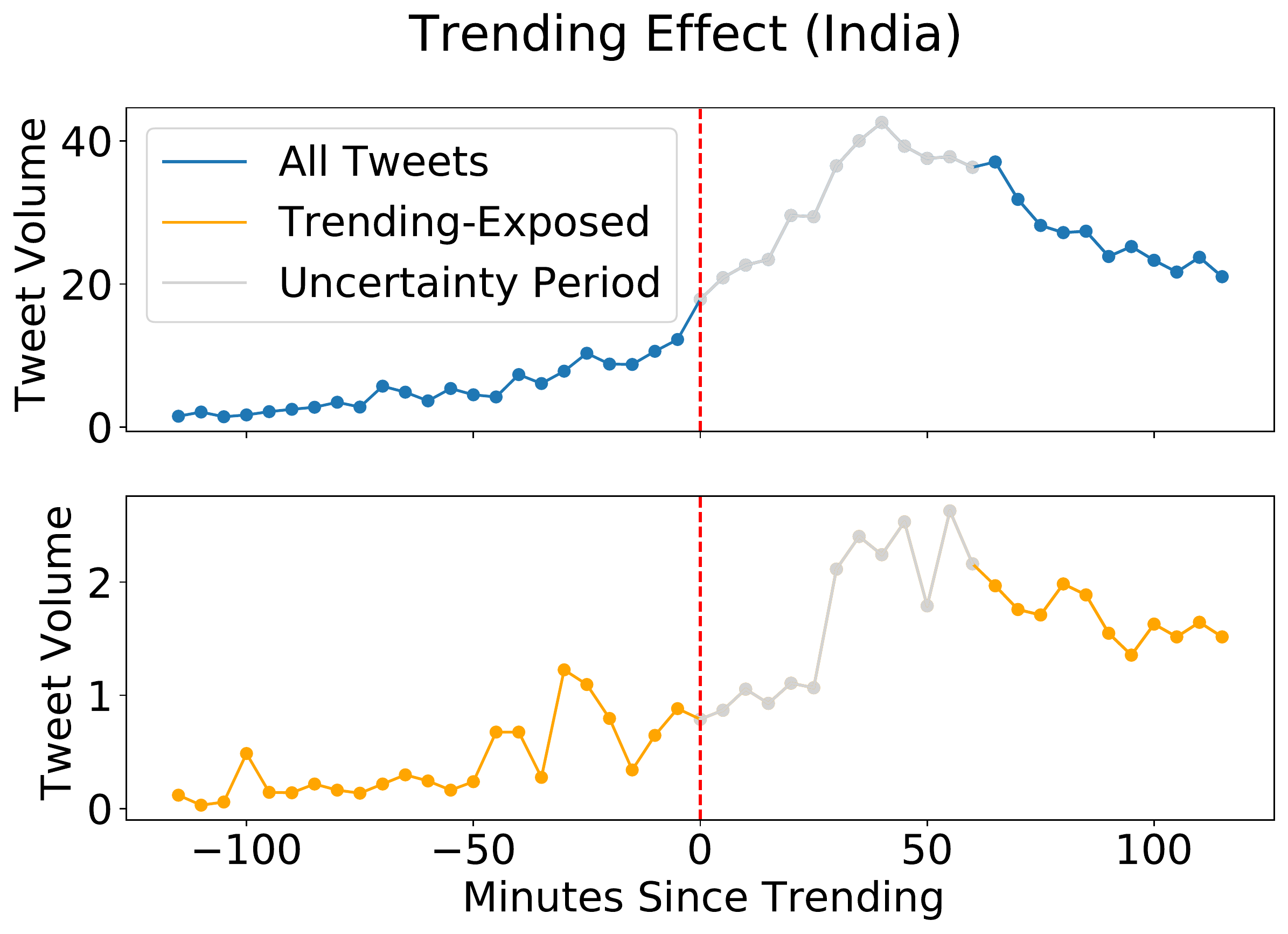}
	\caption[Donut Hole Event Study]{Average number of tweets (for every 5 minute bucket) in the Indian dataset. The top panel shows the average across all the tweets, while the bottom panel shows on only the tweets that we estimate to be potentially exposed through the trending topics page. Time is adjusted to be minutes from appearing on the trending topics page. The figure also shows our donut hole approach to handling trending time uncertainty. In the \texttt{Donut hole} model, we drop tweets in the 60 minute uncertainty period where the treatment status is uncertain.}
	\label{fig:event_studies_india}
\end{figure}

We depict this phenomenon in Figure~\ref{fig:event_studies_india}, which shows the average tweet volume of all the hashtags in our Indian dataset. The x-axis shows the minutes since the hashtag was trending, where the zero indicates the top of hour in which the hashtag was trending.
The figure clearly shows how the \emph{Donut hole} approach might produce an under-estimate since we ignore a large chunk of data and the effects of trending do not last long.

\subsection{Results}
\label{sec:results}

\begin{figure}[ht]
	\centering
	\includegraphics[width=\linewidth]{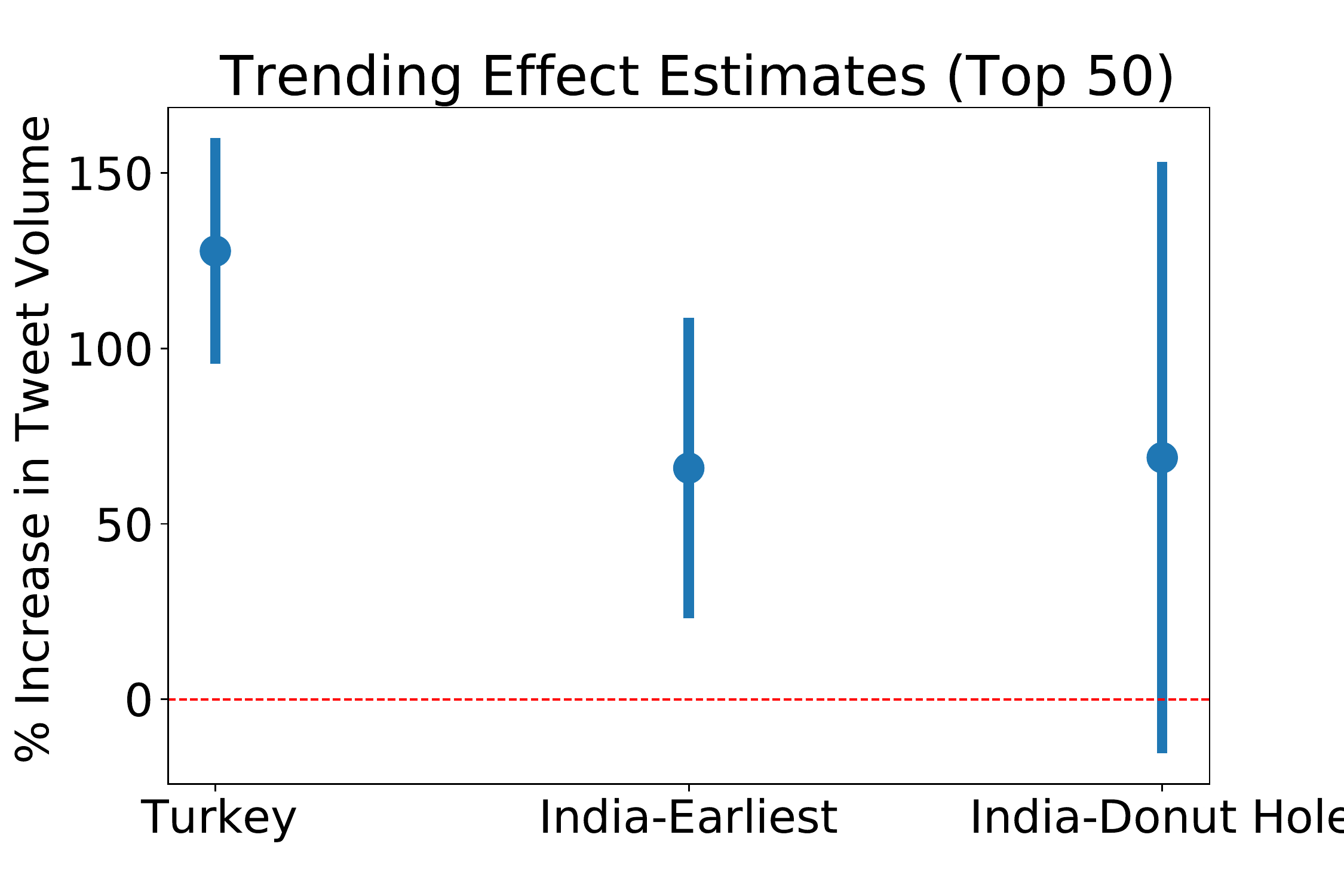}
    \caption[Trending Effect Sizes]{Estimated effects of a hashtag trending in the top 50 for both datasets. Error bars are cluster-robust 95\% confidence intervals clustered on hashtag.}
	\label{fig:trending_effect_coefs_top_50}
\end{figure}

Figure \ref{fig:trending_effect_coefs_top_50} shows the effects of trending for both the datasets. 
The figure shows the percentage increase in the tweet volume.\footnote{We plot the exponent of the regression coefficient along with the 95\% confidence intervals. Since we are using a quasi-Poisson model, the coefficient indicates the effect on a logarithmic scale.}
We find that appearing on the trending topics page leads to an immediate increase in tweet volume of between 60 and 130\% on average.
For India, we also present results for two scenarios: \emph{Earliest} and \emph{Donut hole}. The point estimates remain similar for both approaches, however, as expected, the confidence intervals for the \emph{Donut hole} approach are wider.
The full regression output for both datasets can be found in Table \ref{tab:regression_tables} in the Appendix.

\subsubsection{Effect of trending position.}
The position of the hashtag in the trending list might affect the returns of trending. 
A trending hashtag can be the first item displayed on the trending topics page or the 50$^{th}$.
To understand the impacts of the position, we modify our model in Equation \ref{eq:trending_causal_model} to include an additional effect for when a hashtag reaches the top 10 of the trending topics page, which can result in the topic appearing on the home page, at least for Web users.
This gives the following model:
\begin{equation}
	Y_{th} = \alpha + \lambda D_{th} + \tau t + \gamma (D_{th} \times t) + \beta E_{th} + \rho D^{10}_{th} +  \xi_h + \epsilon_{th}
	\label{eq:trending_causal_model_modified}
\end{equation}
where $D^{10}_{th}$ is a dummy for whether hashtag $h$ is in the top 10 of trends at time $t$, $\rho$ gives the additional return to a hashtag moving from the top 50 to the top 10 and $\rho + \lambda$ can be thought of as the total return to reaching the top 10.
This model relies on a hashtag reaching the top 50 and the top 10 at different times. 
This requires more granularity than the Indian dataset's hourly trending data, and thus this analysis can only be performed on the Turkish dataset, with it's trending data available every 5 minutes.
In Turkey, all the 418 trends that reached the top 50 also reached the top 10 and a total of 89 hashtags entered the top 50 and top 10 at different times.
When we study both treatment effects, we find that trending causes a $ \approx 130\% $ increase in new tweets using the hashtag. 
There is an additional boost of $\approx 4 \% $ when a hashtag enters the top 10.
While this top 10 ``boost'' alone is not statistically significant, the \textit{total} effect of trending in the top 10 is jointly significant ($\lambda + \rho \neq 0$ with $p < .0001$ ).

\subsubsection{Effects over time.}
Next, we look at the returns to trending over time. As can be seen in Figures~\ref{fig:event_studies_turkey} and ~\ref{fig:event_studies_india}, the effects of a hashtag trending appear to dissipate quickly. To quantify this effect, we look at the coefficient of the interaction of trending and time in Equation 1 in Table~\ref{tab:regression_tables}.
While there is a positive slope prior to trending, this is negated or reverses once the hashtag is trending.

\subsubsection{Potential bias and external validity.}
While there is some possibility that some trending exposed tweets were not caused by the trending topics page (they may have been from the algorithmic timeline, false positives from gaps in the friend network (from private users), or artifacts of deleted tweets), which would tend to overstate the trending effect, many of the assumptions we made are conservative.
By construction, any users that were exposed through their follower network cannot be impacted by the trending topics page.
This tends to underestimate a trending effect because it is very possible for a user to not see every tweet that appears on their timeline.
In the Indian case study, for example, many users are only exposed to the hashtag by accounts like BJP politicians and official BJP accounts, who have millions of followers.
As soon as an account like \texttt{@bjp4india} (17.4 million followers in January 2022) uses one of the hashtags, millions of users on the network are thereby network exposed and thus ``disqualified'' --- according to our assumptions --- from contributing to the trending effect.
Our analysis shows that while only 2.5\% of all users belonged to the trending exposed group, $\approx$16\% of users tweeted a hashtag after \textit{exactly one} friend used the hashtag.
It is possible that some of these users with only one network exposure may have actually been caused by trending topics page exposure.
This is not an issue of concern with Turkey because the campaigns were not started by high-centrality users.

In addition to being an underestimate, this estimate is likely specific to these circumstances (artificially seeded Turkish trends or Indian political Twitter and pro-BJP political hashtags).
The trending topics page in different markets covering different subjects likely has a different return.
Even within this case study, different hashtags see very different returns, which is consistent with the idea that content has endogenous virality~\cite{goel2016structural}.

While our estimates may be smaller than the real effect size, the true returns to trending are likely still modest.
We also point to caution in interpreting the results --- a 60 to 130\% increase may sound large, but this should be put in the proper context of the campaigns, which involve very few tweets prior to the campaign beginning and thus prior to them trending.
Much of the hashtag adoption in these campaigns can be explained by organic spread via the follower network.
This suggests that Twitter's trending topics page might not be such a powerful agenda setting tool.
That is, simply (artificially) promoting a trend to the trending topics page does not cause massive widespread adoption.
Consistent with the literature, most of the hashtag adoption comes from the influence by high-degree users and network diffusion.
Still, the trending topics page may be useful to a campaign like the cases used in this paper by enabling the hashtag to reach new audiences outside the range of network diffusion. 
We explore this topic in the following section.

\subsection{Who participates because of exposure through the trending topics page?}
While we quantified the causal effect of a hashtag appearing on the trending topics page, an important unanswered question is \textit{who} is reached by the trending topics page.
One argument that is often made on the benefits of having a trending page is that it helps solve the cold start problem for new users and expose users to content which they might not have seen otherwise.
In this section, we explore the characteristics of the users who participate after a hashtag trends.
Our results, using data from both India and Turkey, show that users exposed through the trending topics page are less popular, less active, lower degree users, but are able to expose more of their followers because they spread the trend in a previously unexposed part of the Twitter network.

\begin{figure}[ht!]
\centering
\begin{minipage}{.55\textwidth}
\centering
\begin{subfigure}{\textwidth}
  \includegraphics[width=\textwidth]{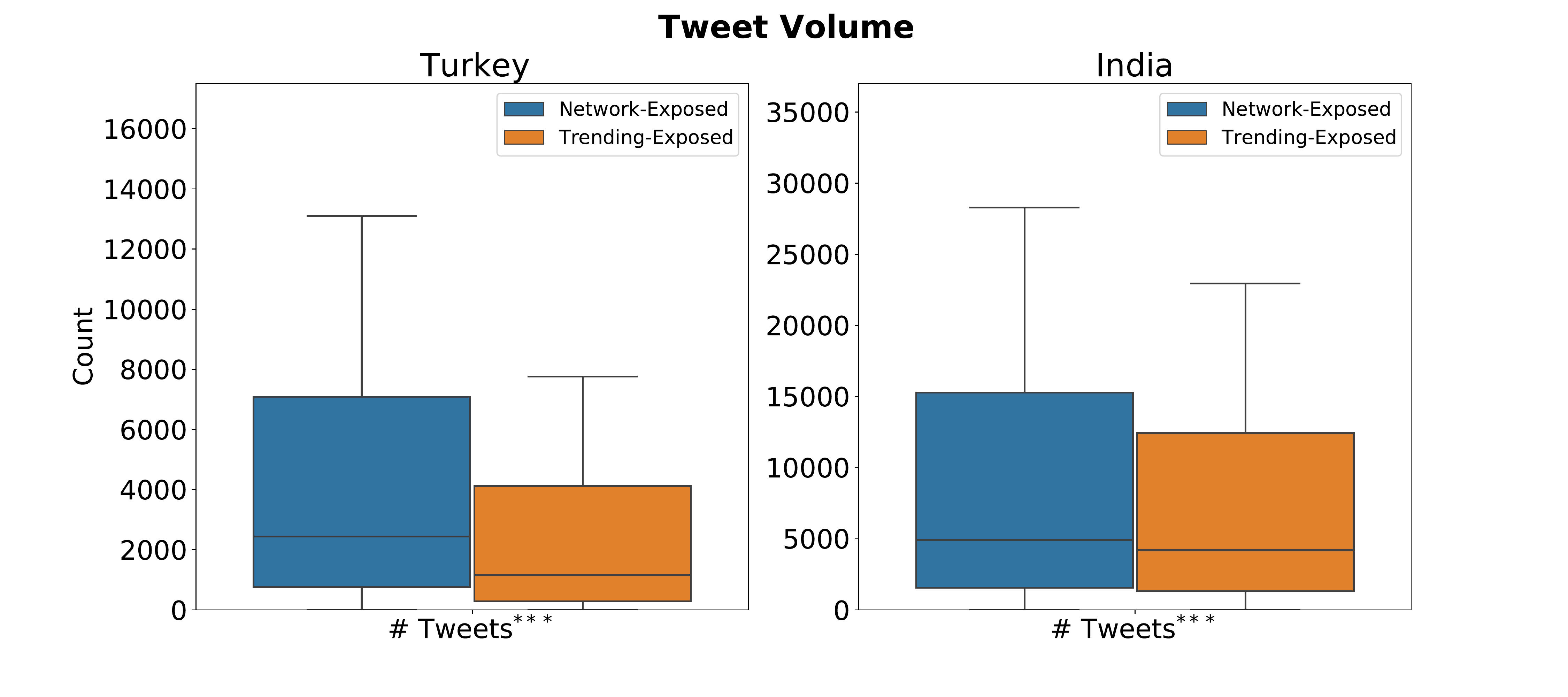}
  \caption{}
\end{subfigure}
\end{minipage}%
\par\medskip
\begin{minipage}{.5\textwidth}
\centering
\begin{subfigure}{\textwidth}
  \includegraphics[width=\textwidth]{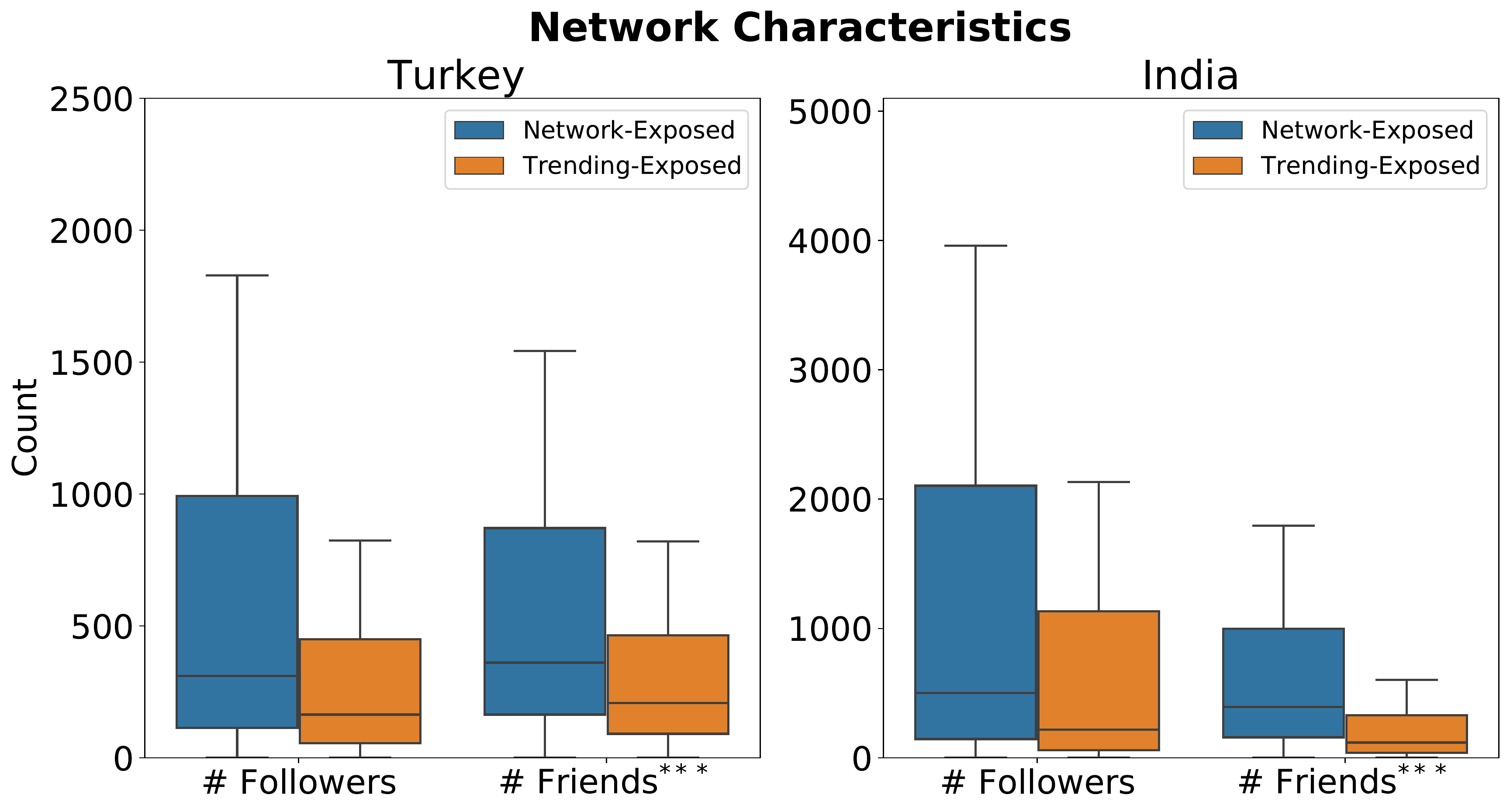}
  \caption{}
\end{subfigure}
\end{minipage}%
\par\medskip
\begin{minipage}{.55\textwidth}
\centering
\begin{subfigure}{\textwidth}
\centering
  \includegraphics[width=0.5\textwidth]{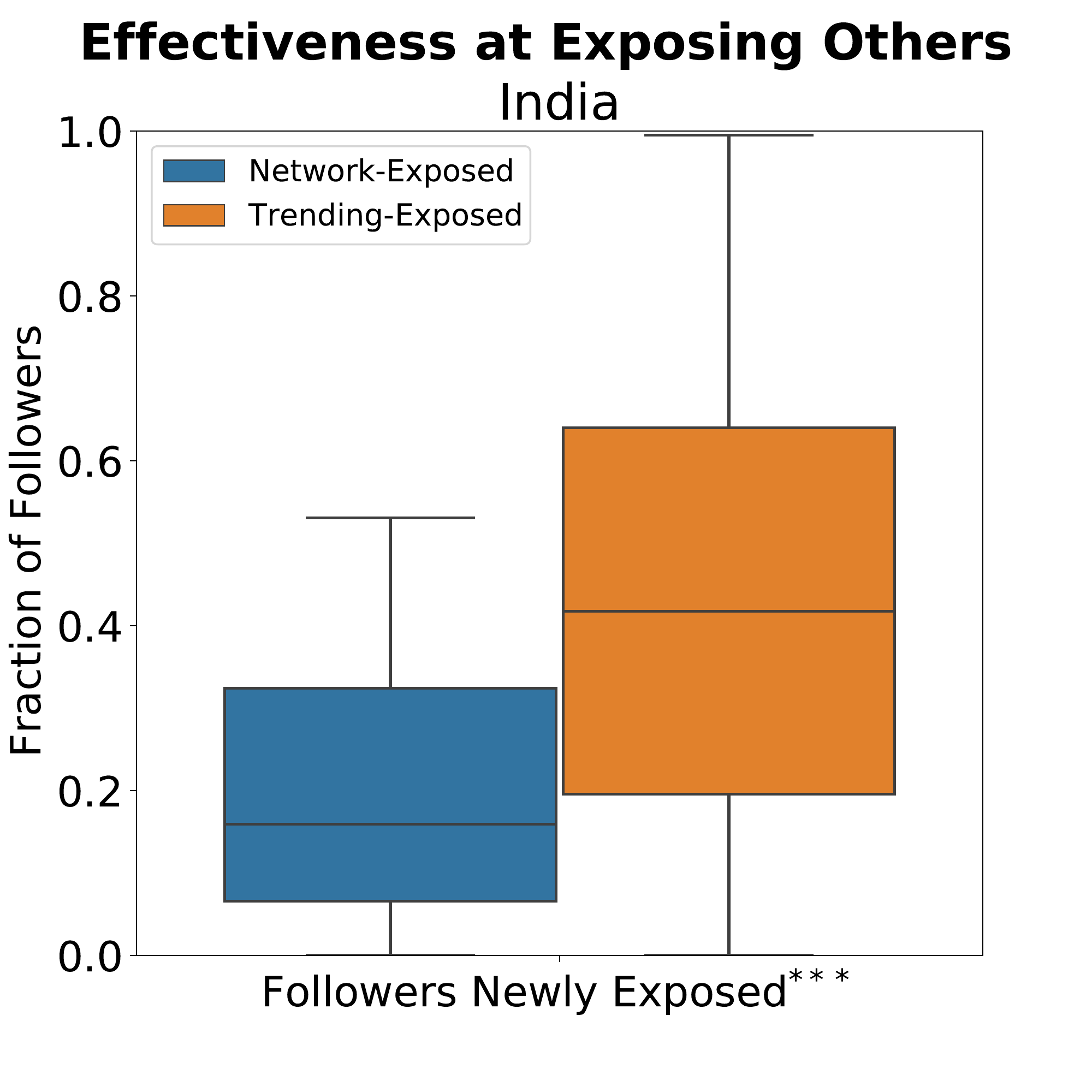}
  \caption{}
\end{subfigure}
\end{minipage}%
\par\medskip
\caption{(a) Tweet volume, (b) Network characteristics, and (c) Effectiveness of exposing others. The characteristics that are statistically significant are marked with ***.}
\label{fig:user_characteristics}
\end{figure}

We compare users across the network exposed and trending exposed groups.
For each group, we compare users along three characteristics: activity (tweet volume), network measures (such as followers/friends), and their effectiveness to convince others.
Figure~\ref{fig:user_characteristics} shows the results.

In both our datasets, participants who are exposed to the trend via trending are generally less active (Figure~\ref{fig:user_characteristics} (a), posting less number of tweets), and less popular (Figure~\ref{fig:user_characteristics} (a), with fewer number of friends) when compared to the users who were exposed through the network. Both these differences are statistically significant at the $p$ \textless .01 level. 
Moreover, those exposed to the trend through the trending topics page are less connected to the main Twitter network, as evidenced by their follower and friend numbers (Figure~\ref{fig:user_characteristics}(b)).

Despite being less popular, these users are still effective at exposing new users to the trend by bringing the hashtag to a relatively unexposed Twitter community (Figure~\ref{fig:user_characteristics} (c)).
To compute the effectiveness of exposing new users, we first identified users who were trending exposed, and then looked at what fraction of their followers tweet a hashtag \textit{after} they tweet the hashtag.
Users reached through the trending topics page often expose a significantly larger fraction of their followers compared to users who were network exposed. On average, while around 40\% of the followers of the trending exposed users were exposed through them, only 20\% of the followers of the network exposed group were reached.
This provides evidence that the trending topics page allows the campaigns to reach new users who are not as connected to pro-BJP Indian political Twitter.\footnote{We only collected the follower data only for India, and hence can provide these estimates only for India.}

In general, these findings support the idea that trending can be important for expanding an idea beyond a highly connected sub-network.
This may be especially important in considering the impact of trends, where manipulated campaigns like the political campaigns in the Indian case reach new audiences.
Spreading the hashtag within pro-BJP Twitter echo chamber may not provide a return on investment in creating the infrastructure to manipulate the trends from an agenda setting standpoint because the users are already sympathetic to the BJP cause.
However, if the trending section helps the hashtag reach a more diverse set of users, this can be quite helpful in setting the agenda in more moderate Twitter communities.

\section{Discussion}
\label{sec:discussion}
In this paper, using carefully curated datasets, we study the causal impact of a hashtag appearing on the trending topics page on Twitter. Our analysis suggests that the returns to trending are limited.
While in both case studies there is a detectable boost in engagement from trending, the effect size is small.
Simply put, the trending topics page cannot create a massively popular hashtag on its own; the spread through the network still plays an important role in creating viral trends.
However, the trending topics page does give an ambitious agenda-setter the opportunity to reach a broader audience than what is just possible through network seeding.

Even with all the steps we took in conducting the analysis and reporting the results, our work has limitations, which mostly arise from data access issues.
First, there is the issue of accessing historical trending information, which is challenging to collect and lacks the granularity desired for this kind of study. Even with our effort to collect trending data in real time, without access to granular trending data, our estimates can have bias and lower statistical precision (particularly in our Indian dataset).
Secondly, our analysis hinges on exposure through the follower network, and we posit that everyone who follows a user is exposed by their content instantly after tweeting. To the extent that this misclassifies users who are induced to use a hashtag as network, rather than trending exposed, we expect this attenuates our estimates. Since exposure data is only available to the social media companies~\cite{lazer2020studying}, this is a systemic problem for any academic research paper trying to audit these social media systems.
Thirdly, we assume that exposure through the trending topics page leads people to tweet the hashtag. People may be influenced by the trend, yet not be inspired to use the hashtag and may, for instance, just retweet or like content with the hashtag.
We have no way of measuring such an effect.

Finally, we lack some important information about algorithms for trending topics and other parts of Twitter. The exact algorithm behind the trending topics section is unknown to external researchers.
There are some exposure sources, such as exposure to content in Twitter's main algorithmically-ranked feed that is not from accounts followed by a user (e.g. the ``liked by @user'' feature), which we have not accounted for here.
The amount of such content that is included in users' feeds (starting in 2017) has changed over the years, thereby perhaps modifying both effects of trending topics and our ability to distinguish this from other exposures~\cite{koumchatzky2017using}. 
That is, as is the issue with any audit study, the exact technology used is constantly evolving.
To check whether algorithmic ranking influences how trending hashtags are shown to users, we obtained data collected by \citet{bandy2021curating} to compare chronological and non-chronological (i.e. algorithmic) ranking. From the 3,000 hashtags from their data, we looked for hashtags which were trending on the day the data was collected which gave us 148 hashtags.
For these hashtags, we looked at whether their ranking in the feed is significantly different in the algorithmic vs. chronological timeline.
We find no significant difference (p $\textgreater$ 0.1).
Even though this does not provide concrete evidence that trending hashtags were not impacted by algorithmic ranking in the cases we study, we could build on such measurement to study the effects of algorithmic ranking in future work.

Despite these limitations, we believe our study makes an impactful contribution to the literature. Even a lower bound, context specific result is interesting, since our methodology is generic and can be extended to include other cases where astroturfing happens. 
We also use multiple large datasets from different real world contexts. The datasets, by design can partly help address some of the limitations. For instance, in the Turkish dataset, all the tweets that are used to make the hashtag trend are deleted in a short time frame. So the chances of users getting exposed through the network (at least in the initial few minutes) is quite low.

\subsection*{Ethics Statement}
The paper provides analysis and results on a highly relevant and current topic: the ethics of algorithmic amplification.
Our paper provides concrete, data driven evidence to build the case for a debate on trending topics.
As we saw from our results, the effects of trending might be short lived but depending on the case, they might be impactful.
As with the two cases used in our analysis, Twitter trends have been widely manipulated and artificially manipulated trends happen all the time.
There have been numerous calls to disable trending topics, due to posited harms and their susceptibility to manipulation.\footnote{e.g. See \url{https://bit.ly/33EuwK8}}
In many cases, the manipulation of trending topics is readily detectable and can be fixed easily, though perhaps such campaigns would effectively adapt in response to such efforts..
The Indian WhatsApp campaigns are easily detectable based on their repetitive content \cite{jakeschTrendAlertHow}.
With the Turkish ephemeral trend attacks, Twitter could simply account for deleted tweets in the trending algorithm \cite{elmasLateralAstroturfingAttacks2019}.
In October 2020, in preparation for the 2020 US election, Twitter started adding more context to their trending topics \textit{just} for users in the United States, thus manually curating trending topics. Compared with the U.S., where trends are at most rarely astroturfed, astroturfed trending topics are apparently more common in the Global South, in countries such as Nigeria, India, and Brazil. This can reflect a pattern of U.S. tech companies neglecting less lucrative foreign markets.
This paper contributes evidence about the prevalence and consequences of trending topics, which could feature in arguments for greater effort ensuring the integrity of trending topics or removing that feature where they cannot.

\section{Acknowledgements}

J.S. was funded by the MIT Lincoln Laboratory Military Fellowship.
K.G. completed part of this work while at MIT and was supported by a Michael Hammer postdoctoral fellowship. His research is supported through funding from Knight Foundation, Google and NSF.
D.E. was a paid consultant to Twitter while this work was undergoing revision. K.G. and D.E. have received received funding for other work from Meta Platforms.

\section*{Appendix}

\subsection*{Regression Tables}
The full regression results are shown in Table~\ref{tab:regression_tables}.

\begin{table*}
\centering
\caption{Regression results}
\label{tab:regression_tables}
\renewcommand*{\arraystretch}{1.3}
\begin{tabular}{lcc|cc} 
\hline\hline
                            & \multicolumn{2}{c|}{Turkey}                 & \multicolumn{2}{c}{India}                                         \\ 
\hline
                            & Combined             & Standard                & Earliest               & Donut Hole                                     \\
\hline
Intercept                   & 2.654$^{***}$          & 2.778$^{***}$          & 1.282$^{***}$          & 1.406$^{***}$                                \\
                            & (0.159)              & (0.153)              & (0.155)              & (0.144)                                    \\
Trending ($\lambda$)        & 0.836$^{**}$           & 0.823$^{***}$          & 0.506$^{***}$          & 0.524$^*$                                  \\
                            & (0.224)              & (0.142)              & (0.182)              & (0.312)                                    \\
Exposed Tweets ($\beta$)    & 0.004$^{***}$          & 0.003$^{***}$          & 0.033$^{***}$          & 0.034$^{***}$                                \\
                            & (0.001)              & (0.001)              & (0.006)              & (0.007)                                    \\
Time ($\tau$)               & 0.016$^{***}$          & 0.027$^{***}$          & 0.007$^{***}$          & 0.007$^{***}$                                \\
                            & (0.008)              & (0.007)              & (0.003)              & (0.003)                                    \\
    Time $\times$ Trending ($\gamma$)  & -0.019$^{**}$          & -0.027$^{***}$         & -0.007$^{***}$         & -0.007$^*$                                 \\
                            & (0.008)              & (0.007)              & (0.003)              & (0.004)                                    \\ 
Trending in Top 10 ($\rho$) & 0.042                 &                      &                      &                                            \\
                            & (0.245)              &                      &                      &                                            \\
\hline

Observations                & 10,327               & 7,346                & 2,441                & 1,790                                      \\
$R^2$                       & 0.856                & 0.855                & 0.691                & 0.698                                      \\ 
\hline\hline \\[-1.8ex]
\textit{Note:} & &  \multicolumn{3}{r}{$^{*}$p$<$0.1; $^{**}$p$<$0.05; $^{***}$p$<$0.01} \\
& & \multicolumn{3}{r}\textit{Standard errors clustered at hashtag level.} \\
\end{tabular}
\vspace{-\baselineskip}
\end{table*}

\bibliography{main-no-zotero-sync}
\end{document}